\documentclass[manuscript]{aastex}
\usepackage{amssymb}
\usepackage{verbatim}

 \pdfoutput=1
 
\shorttitle{Damping of sound waves} \shortauthors{Ib\'{a}\~{n}ez et al.}

\begin{document}

\title{Damping of sound waves by bulk viscosity in reacting gases}
\author{Miguel H. Ib\'{a}\~{n}ez S.\altaffilmark{1}}
\author{Pedro L. Contreras E.}
\affil{Centro de F\'{\i}sica Fundamental, Universidad de los Andes,
M\'erida 5101, Venezuela.}
\email{pcontreras@ula.ve}

\altaffiltext{1}{Now at Valle de Villa de Leyva, Colombia.}

\begin{abstract}
The very long standing problem of sound waves propagation in fluids is
reexamined. In particular, from the analysis of the wave damping in reacting
gases following the work of Einsten \citep{Ein}, it is found that the damping due to
the chemical reactions occurs nonetheless the second (bulk) viscosity
introduced by Landau \& Lifshitz \citep{LL86} is zero. The simple but important case of a
recombining Hydrogen plasma is examined.
\end{abstract}

\maketitle

\section{Introduction}

Propagation of disturbances, in particular sound waves in hypothetical
equilibrium fluids has been researched since the pioneer works
\citep{Raly,Ein,Lam,Lins,Lig} and their main characteristics have been well
established, i.e. waves propagate with certain velocity, and are damped by
the irreversible processes say viscosities, thermal conduction and chemical
reactions. Landau \& Lifshitz \citep{LL86} introduce a bulk (second) viscosity coefficient $%
\zeta$ in the equation of motion for accounting the dissipation of energy
due to compression or expansion through transferring kinetic energy into
internal degrees of freedom (such as chemical reactions, excitation of
atomic/molecular levels, etc.). However, in the case of chemical reactions
such approximation only holds if one neglects any others effects except the
density change $\delta\rho$ due to the chemical reaction.

Henceforth, as it will be shown at the present note, the Landau
approximation is rather restrictive. In fact, if $\xi$ is a parameter
characterizing the degree of advance of chemical reaction in the fluid (say,
the concentration of one chemical component) and $\xi_{0}$ its respective
value at chemical equilibrium, which generally is a function of the
equilibrium density and temperature, say $\xi_{0}(\rho_{0},T_{0})$ \citep{VK}%
; henceforth, as it can be realized, in the Landau approximation %
\citep{LL86,Ib09} the second viscosity coefficient is $\sim(\partial\xi_{0}/%
\partial\rho)_{T}$, therefore, when $(\partial\xi_{0}/\partial\rho)_{T}=0$
the acoustic wave damping is also zero. However, when $(\partial\xi_{0}/%
\partial T)_{\rho}\neq0$ the sound waves could be damped nonetheless the
Landau bulk viscosity coefficient is zero, as it will be shown below.

The present analysis on the bulk viscosity is made for any reacting gas
where the chemical reactions can be reduced to a net reaction that can be
described by one parameter measuring the advance of the reaction
\citep{Yo73,IP83}. However, for context, the results are applied to a
Hydrogen plasma where the simple reaction $H^{+}+e^{-}$ $\rightleftarrows$ $%
H+(\chi)$ proceeds ($\chi$ being the ionization potential). The knowledge of
the above plasma is of particular importance in Astrophysics, say, the solar
atmosphere \citep{SS72,SL74,S78,Boh87,NU90}, the interstellar gas %
\citep{S78,S82,S90} and more recently in the Intracluster gas %
\citep{FS03,RB04,FR05,FF09}, in particular due to the fact that wave
dissipation have been invoked as one of the mechanisms of heat input.
However, a detailed study of the thermal behavior of the above plasmas is
out the scope of the present study, which is particulary restricted to find
an expression of the bulk viscosity coefficient in chemically active plasmas.

\section{Basic Equations}

In general, for a 1-D plane wave the wave number $k$ and the frequency $%
\omega $ are related by
\begin{equation}
k=\frac{\omega }{c}.  \label{k0}
\end{equation}%
The parameter $c$ is defined by the relation
\begin{equation}
c=\pm \sqrt{\frac{\partial p}{\partial \rho }},  \label{c}
\end{equation}%
where
\begin{equation}
\frac{\partial p}{\partial \rho }=\frac{1}{\rho _{0}}\left[ p_{0}-\left(
\frac{\partial (pV)}{\partial V}\right) \right] ,  \label{pro0}
\end{equation}%
with $V=1/\rho $ and the equilibrium values denoted with the subindex $_{0}$%
. The relation (\ref{k0}) formally obtained by non-dispersive media also
holds for dispersive media for which $c$ is a complex quantity (as well as $%
k $) \citep{LL86} and only for disturbances propagating in a non-reacting
ideal fluid becomes the adiabatic sound speed
\begin{equation}
c=c_{s}=\sqrt{\left( \frac{\partial p}{\partial \rho }\right) _{s}}.
\label{cs}
\end{equation}

Strictly speaking, the basic gas dynamic equations admit solutions in the
form $\sim\exp(\beta t+i\mathbf{k.r})$, where $\beta=\sigma-i\omega$ and $%
\mathbf{k=k}_{r}+i\mathbf{k}_{i}$, where $\sigma$ and $\omega$ are real
quantities and $\mathbf{k}_{r}$ and $\mathbf{k}_{i}$ are real vectors.
Therefore, one may write the sound disturbance as $\sim$ $\exp(\sigma
t-k_{i}x)\exp\left[ i\left( k_{r}x-\omega t\right)\right]$ for the
one-dimensional problem. The above can be interpreted as a wave of frequency
$\omega$, wavelength $\lambda=2\pi/k_{r}$, traveling along the \emph{x-axis}
with a phase velocity $v=\omega/k_{r}$ and the amplitude $\sim\exp(\sigma
t)\exp(-k_{i}x)$, the first factor measures the attenuation (or growth if $%
\sigma>0$) in time, and the second factor measures the spatial absorption
(or amplification if $k_{i}<0$) in ordinary progressive wave propagation
studies \citep{MBL51,LL81,IM81}. The present analysis is restricted to the
spatial absorption of linear wave propagation in a chemically active fluids
from where the bulk viscosity coefficient is calculated.

For reacting gases if the set of "chemical reactions" which are in progress
can be reduced to a single reaction $\sum_{j}\nu_{j}A_{j}=0$ where $A_{j}$
are the chemical symbols of the reagents and the coefficients $\nu_{j}$ are
positive or negative integers, there is at least on component $j$ for which
the concentration $\xi_{j}=n_{j}/n$ goes to zero when the reaction proceeds
to a sense indefinitely, here $n$ denotes the total number
density of atoms and $n_{j}$ is the number density for gas particles of the $%
j-th$ component. So, one may introduce the parameters $\xi$, and $a$, such
that

\begin{equation}
\xi _{j}=\frac{n_{j}}{n}= a \; \xi, \qquad 0\leq \xi \leq 1;  \label{xi}
\end{equation}
where $\xi$ denotes the degree of advance of the reaction and $a$ the
maximum number of abundance ratio of the $j-th$ component to the total
number of nuclei.

From the equation of continuity for the different components and the
definition (\ref{xi}) one may obtain the rate equation \citep{VK,Yo73,IP83}
\begin{equation}
\frac{d\xi}{dt}+X(\rho,T,\xi(\rho,T))=0,  \label{X0}
\end{equation}
where $X(\rho,T,\xi(\rho,T))$ is the net rate which at equilibrium $X(\rho
_{0},T_{0},\xi_{0}(\rho_{0},T_{0}))=0$.

Additionally, an ideal-like state equation will be assumed, i.e.
\begin{equation}
p=\frac{R \; \rho \; T}{\mu(\xi)},  \label{p0}
\end{equation}
where $R$ is the gas the gas constant constant and $\mu(\xi)$ is the mean
molecular weight, $\mu^{-1}=\sum_{j}\xi_{j}$.

On the other hand the internal energy per unit mass becomes
\begin{equation}
u=A(\xi )RT+\chi N_{0}a\xi ,  \label{energy}
\end{equation}%
where $\chi $ and $N_{0}$ denote the dissociation energy and the Avogadro's
number and
\begin{equation}
A(\xi )=\sum_{j}\frac{\xi _{j}}{\gamma _{j-1}},  \label{A}
\end{equation}%
$\gamma _{j}$ being the specific heat-ratio for the $j-th$ component.

For an adiabatic change, the energy equation can be written as
\begin{equation}
RA(\xi )\frac{\delta T}{\delta t}-\frac{p}{\rho ^{2}}\frac{\delta \rho }{%
\delta t}+RTB(\xi ,T)\frac{\delta \xi }{\delta t}=0,  \label{Ener0}
\end{equation}%
where
\begin{equation}
B(\xi ,T)=\frac{dA}{d\xi }+\frac{a\chi }{k_{B}T},  \label{BB}
\end{equation}%
being $k_{B}$ the Boltzmann constant. For linear disturbances close to the
equilibrium
\begin{equation}
RA_{0}\delta T-\frac{p_{0}}{\rho _{0}^{2}}\delta \rho +RB_{0}T_{0}\delta \xi
=0,  \label{adiaba}
\end{equation}%
where $A_{0}=A(\xi _{0})$ and $B_{0}=B(\xi _{0},T_{0})$ are the equilibrium
values of the functions $A(\xi )$ and $B(\xi ,T)$.

For fluctuations $\sim\exp(-i\omega t)$ from Eq.(\ref{X0}) follows that the
disturbances $\delta\xi$, $\delta\rho$, and $\delta T$ are related by the
equation
\begin{equation}
\delta\xi=\frac{\xi_{\rho}^{\ast}}{1-i\omega\tau}\delta\rho+\frac{\xi
_{T}^{\ast}}{1-i\omega\tau}\delta T,  \label{delxi}
\end{equation}
where $\tau=(\partial X/\partial\xi)^{-1}$ is the relaxation time which is a
positive quantity for chemically stable gases; and where $\xi_{\rho}^{\ast
}=\left( \partial\xi_{0}/\partial\rho\right) _{T}$, and $\xi_{T}^{\ast}=%
\left(\partial\xi_{0}/\partial T\right) _{\rho}$ are the derivatives at
equilibrium \citep{Yo73,Ib,IP83}.

Additionally, from Eqs.(\ref{p0}), (\ref{X0}),and (\ref{adiaba}) the Eq.(\ref
{pro0}) becomes

\begin{equation}
\frac{\partial p}{\partial \rho }=\frac{p_{0}}{\rho _{0}}\left[ 1+Q\right] ,
\label{Ein0}
\end{equation}%
the $Q$ factor is given by
\begin{equation}
Q=\frac{\left( 1-i\omega \tau \right) -\left( \mu _{0}B_{0}+\mu _{\xi
}A\right) \rho \xi _{\rho }^{\ast }\,-T\xi _{T}^{\ast }\mu _{\xi }/\mu }{\mu
_{0}\left[ A\left( 1-i\omega \tau \right) +T\xi _{T}^{\ast }\,B_{0}\right] },
\label{Q}
\end{equation}
$\mu _{\xi }$ being the derivative of the molecular weight with respect to
the chemical parameter. It is important to mention that the above relation (%
\ref{Ein0}) for a particular simply chemical reaction was obtained in an
early paper by \citep{Ein}.

In the limiting when $\omega \tau \rightarrow \infty $ (frozen chemistry), $%
Q\rightarrow 1/A\mu _{0}$ and in the opposite limiting $\omega \;\tau
\rightarrow 0$ (the chemical equilibrium follows the fluctuation)
\begin{equation}
Q=\frac{1-\left( \mu _{0}B_{0}+\mu _{\xi }\,\,A\right) \rho \xi _{\rho
}^{\ast }\,-T\xi _{T}^{\ast }\mu _{\xi }/\mu }{\mu _{0}\left[ A+T\xi
_{T}^{\ast }B_{0}\right] }.  \label{Q1}
\end{equation}

In the limiting case when the fluctuation $\delta \xi $ is only due to the
change of density $\xi _{T}^{\ast }=0$, the Eq.(\ref{Q}) reduces to
\begin{equation}
Q=\frac{\left( 1-i\omega \tau \right) -\left( \mu _{0}B_{0}+\mu _{\xi
}A\right) \rho \xi _{\rho }^{\ast }}{\mu _{0}A\left( 1-i\omega \tau \right) }%
.  \label{xiT0}
\end{equation}

On the opposite limit when $\xi _{\rho }^{\ast }=0$,
\begin{equation}
Q=\frac{\left( 1-i\omega \tau \right) -T\xi _{T}^{\ast }\mu _{\xi }/\mu }{%
\mu _{0}\left[ A\left( 1-i\omega \tau \right) +T\xi _{T}^{\ast }B_{0}\right]
}.  \label{xir0}
\end{equation}

If in the limiting $\xi_{T}^{\ast}=0$ additionally $\xi_{\rho}^{\ast}=0$,
henceforth $Q=1/\mu_{0}A=\gamma-1$ and therefore from Eq.(\ref{Ein0}) $\sqrt{
\partial p/\partial\rho}$ $=$ $\sqrt{\gamma p_{0}/\rho_{0}}$ (being $\gamma$
the specific heat ratio) becomes the isentropic sound speed $c_{s}^{2}$ in a
non-reacting ideal gas, as it should be.

It is  interesting to point out that in the Landau approximation Landau \& Lifshitz \citep{LL86}
$(pp. \; 308 - 312)$, where the fluctuation $\delta \xi $ is assumed to occur at
constant entropy $S$, i.e. the change of pressure $p$ is due only to the
change of density $\delta \rho $ produced by the fluctuation in the chemical
parameter $\delta \xi $,
\begin{equation}
\frac{\partial p}{\partial \rho }=\frac{1}{1-i\omega \tau }\left[
c_{0}^{2}-i\omega \tau c_{\infty }^{2}\right] ,  \label{proLL}
\end{equation}%
and $c_{0}$ is given by
\begin{equation}
c_{0}^{2}=\left( \frac{\partial p}{\partial \rho }\right) _{eq}=\left( \frac{%
\partial p}{\partial \rho }\right) _{\xi }+\left( \frac{\partial p}{\partial
\xi }\right) _{\rho }\left( \frac{\partial \xi _{0}}{\partial \rho }\right)
,\qquad c_{\infty }^{2}=\left( \frac{\partial p}{\partial \rho }\right)
_{\xi }.  \label{ccLL}
\end{equation}

From Eq. \ref{Q} one obtain the corresponding
parameter $Q_{L}$ in the Landau approximation, i.e.

\begin{equation}
Q_{L}=-\frac{1}{1-i\omega \tau }\frac{\mu _{\xi }}{\mu _{0}}\rho \xi _{\rho
}^{\ast }.  \label{QL}
\end{equation}

Finally, in the limiting case when $\xi _{\rho }^{\ast }=0$, it follows that
$\sqrt{\partial p/\partial \rho }$ $=\sqrt{p_{0}/\rho _{0}}$, i.e. the sound
propagation would occur with the isothermal sound speed as it is expected.
Additionally, at the Landau's approximation the effects of the chemical
reaction may be accounted for introducing a second viscosity coefficient in
the motion equation given by the following expression
\begin{equation}
\zeta =\frac{\rho _{0}\tau }{1-i\omega \tau }\left[ c_{\infty }^{2}-c_{0}^{2}%
\right] =\frac{\rho _{0}\tau }{1-i\omega \tau }\frac{p_{0}\mu _{\xi }}{\mu
_{0}}\xi _{\rho }^{\ast },  \label{LLvis2}
\end{equation}
i.e. the Landau bulk viscosity coefficient (in $g$ $cm^{-1}s^{-1}$), as it
can be readily verified from Eq.(\ref{ccLL}).

\section{Collisionally ionized Hydrogen plasma}

\label{appl}

For context, at the present section the above results will
be applied to the simple but important examples of an ionized Hydrogen gas
when it is collisionally ionized. As it will shown the damping of sound
waves becoms zero at the Landau approximation, but different from zero at
Einstein approximation.

A collisionally ionized Hydrogen plasma can be considered as a reacting
plasma where the reaction
\begin{equation}
H^{+}+e^{-}\leftrightarrows H^{0}+\chi ,  \label{colH}
\end{equation}%
proceeds with the following expressions
\begin{equation}
A=\frac{1}{\left( \gamma -1\right) \mu },\;B_{0}=\frac{1}{\gamma -1}+\frac{%
\chi }{k_{B}T},\qquad \mu =\frac{1}{1+\xi },  \label{PhotGas}
\end{equation}
$\xi $ being the degree of the ionization, $\chi $ the Hydrogen ionization
potential and $k_{B}$ the Boltzmann constant, the sub-index $0$ indicating
equilibrium values has been omitted. Additionally, the generalized
ionization recombination rate function \citep{Yo73,IP83} becomes equal to
\begin{equation}
X=N_{0}\rho \alpha (T)\xi ^{2}-N_{0}\rho q(T)\xi (1-\xi )=0,  \label{XCol}
\end{equation}
therefore at equilibrium
\begin{equation}
\xi ^{\ast }(T)=\frac{q(T)}{\alpha (T)+q(T)},  \label{xi0C}
\end{equation}%
the total recombination coefficient $\alpha (T)$ is given by
\begin{equation}
\alpha (T)=\frac{2.06\times 10^{-11}}{T}\left( 0.5\ln \Theta +\frac{0.47}{%
\Theta ^{1/3}}-0.32\right) \frac{cm^{3}}{s},  \label{alfa(T)}
\end{equation}%
and the collisional ionization rate follows the expression
\begin{equation}
q(T)=5.85\times 10^{-11}\sqrt{T}\exp (-\Theta ),\frac{cm^{3}}{s}
\label{q(T)}
\end{equation}
where $\Theta =1.579\times 10^{5}/T$ \citep{Sea59,HSe63,Hum63}. The above
approximation holds \citep{Parker57,CF95} in the range of $3.5\times
10^{3}\lesssim T(K)\lesssim 1.58\times 10^{5}$.

For a collisionally ionized Hydrogen plasma from Eq.(\ref{xi0C}) follows
that $\partial\xi_{0}^{\ast}/\partial\rho=0$, and therefore the second
viscosity in the Landau approximation, Eq.(\ref{LLvis2}), is also zero.
However from Eq.(\ref{Ein0}) the speed of sound $c$ becomes
\begin{equation}
c=\sqrt{\frac{p}{\rho}\left(1+ Q\right)},  \label{ccol}
\end{equation}
with
\begin{equation}
Q=\frac{\left( \gamma-1\right) \left[1 - \mu T\xi_{T}/\left(
1-i\omega\tau\right) \right] }{\left[1+\left(\gamma-1\right) B\mu
T\xi_{T}/(1-i\omega\tau)\right] },  \label{QEins}
\end{equation}
i.e. damping effect occurs due to the irreversible process inherent to the
chemical reaction, as it follows from the fact than $c$ becomes a complex
quantity as well as the wave number $k$ Eq.(\ref{k0}), and which can be
written as $k = k_{r} + i \; k_{i}$, where $k_{r}$, and $k_{i}$ are real
quantities, $k_{i}$ being the damping coefficient which has to be a positive
quantity \citep{Ib}.

On the other hand, from Eqs. (\ref{XCol})-(\ref{q(T)}) the relaxation time
becomes
\begin{equation}
\tau=\frac{1}{N_{0} \; \rho\; q\left(T\right)}.  \label{tau}
\end{equation}
i.e. the relaxation time $\tau>0$ and therefore the Hydrogen plasma is
chemically stable.

The damping per unit wave length $2 \pi k_{i}/k_{r}$ and the phase velocity $%
v_{ph}/c_{T}$ normalized to the isothermal sound speed $c_{T}$ $=\sqrt{%
p_{0}/\rho_{0}}$ have been plotted in Figs. (1a) and (1b) respectively, as
functions of temperature for three different values of $\omega \tau$ ($%
10^{-1}$ dash line, $1$ thick line, and $10$ point line). Regardless the
value of $\omega \tau$ the damping shows maxima, and the phase speed
shows minima at a temperature close to $\log T = 4.16$, temperature at which
the function $\xi_{T}^{\ast}$ becomes a maximum and the effect of the
recombination-ionization process become important. At very low (neutral
Hydrogen) as well as at very high temperatures (ionized Hydrogen), the
damping tends to be zero Fig.(1a), and the sound velocity tends to be the
isentropic sound speed Fig.(1b), as it is expected from simple physical
considerations.

In Figs. (1c) and (1d) the damping per unit wave length ($2\pi k_{i}/k_{r}$)
and the normalized phase velocity ($v_{ph}/c_{T}$) are respectively plotted
but as functions of $\omega \tau $ for temperatures slight lower ($\log
T=4.04$, (dash line) and higher $\log T=4.28$, (point line) than $\log
T=4.16 $ (thick line). The damping per unit wave length becomes a maximum
very close to the $\omega \tau $ value Fig. (1c) where the inflexion point
of $v_{ph}/c_{T}$ occurs as in Fig. (1d). Regardless of the temperature
value, waves with $\omega \tau \ll 1$ propagate as adiabatic disturbances
in a gas at chemical equilibrium, and those with $\omega \tau \gg 1$ as
adiabatic disturbances in a gas where the chemical reaction is frozen. In
the above limiting cases the disturbances tend to be undamped waves Fig.
(1c) as it should be.

\subsection{Photo-ionized Hydrogen plasma}

In this section the results of Section 1 will be applied to a photo-ionized
Hydrogen plasma model i.e. an optically thin Hydrogen plasma ionized by a
background radiation field of averaged photon energy $E$ and
photo-ionization rate $\varsigma $. The net rate function $X(\rho ,T,\xi )$
is given by \citep{CF95} as
\begin{equation}
X(\rho ,T,\xi )=N_{0}\rho \lbrack \xi ^{2}\alpha -(1-\xi )\xi q]-(1-\xi
)(1+\phi )\varsigma,  \label{CFX}
\end{equation}
$\alpha $ is the total recombination coefficient ($cm^{3}s^{-1}$) which is
given be Eq.(\ref{alfa(T)}), $q$ is the collisional ionization rate
($cm^{3}s^{-1}$) according to \citep{Bla81}, $\phi (E,\xi )$ is the number of
secondary electrons which in general is a function of \textbf{the energy
mean photon energy} $E$ and the ionization $\xi $, \citep{Shull85}, and $
\varsigma $ is the photo-ionization rate in $s^{-1}$\citep{Bla81}.
The last term of the right hand side of (\ref{CFX}) just accoaunted for this
effect.Therefore, the corresponding terms in the energy equation
Eq. (\ref{Ener0}), have to be be added for consistency. For accounted the heat
input and output of energy by radiation. So, instead of Equation (\ref{Ener0}) one obtains

\begin{equation}
RA(\xi )\delta T-\frac{p}{\rho ^{2}}\delta \rho +RB(\xi ,T)T\delta \xi
+\delta L\left( \rho ,T,\xi \right) =0,
\end{equation}
Where the net heat/cooling function becomes
\[
L\left( \rho ,T,\xi \right) =N_{0}\{\rho ^{2}\left[ \left( 1-\xi \right) \xi
\Lambda _{eH}\right] +\xi ^{2}\Lambda _{eH^{+}}\}-N_{0}\{(1-\xi
)[E_{h}+(1+\phi )\chi ]\}
\]
where $\Lambda _{eH}$ and $\Lambda _{eH^{+}}$ are
the cooling losses by $e-H$ and $e-H^{+}$ collisions
\citep{Ib} neglecting secondary electrons $\phi =0$

According to \citep{Shull85}, $0.002\lesssim \phi \lesssim $\ $0.366$, for $%
0.95\gtrsim \xi \gtrsim 10^{-4}$ the exact value depending on $E$ (which
depends on the particular optical depth in the gas) strictly speaking a
self-consistent radiative transfer problem should be worked out, and which
is out the scope of the present paper, whose aim is restricted to obtain an
indicative value of the bulk viscosity for a photo-ionized Hydrogen plasma.
Therefore, if in a first approximation the production of secondary electrons
is neglected ($\phi =0$), from Eq. (\ref{CFX}) an explicit form the
ionization $\xi ^{\ast }(\rho ,T)$ at equilibrium can be obtained, i.e.
\begin{equation}
\xi ^{\ast }(\rho ,T)=\frac{N_{0}\rho q-\varsigma +\sqrt{B_{p}}}{2N_{0}\rho
\left( \alpha +q\right) },  \label{xe}
\end{equation}
with
\begin{equation}
B_{p}=\left( N_{0}\rho q+\varsigma \right) ^{2}+4N_{0}\rho \alpha \varsigma ,
\label{B}
\end{equation}
otherwise the solution for $\xi $ at equilibrium becomes an implicit
function of $T$ and $\rho $, and for its calculation one must proceed
numerically. The correction introduced by the secondary electrons is
equivalent to an increase of the value of the photo-ionization rate, as it
can be verified from Eq. (\ref{CFX})

Therefore, from Eq.(\ref{xe}) one obtains
\begin{equation}
\xi _{\rho }^{\ast }=\frac{\rho B_{p\rho }+2\sqrt{B_{p}}\varsigma -2B_{p}}{4%
\sqrt{B_{p}}N_{0}\rho ^{2}\left( \alpha +q\right) },  \label{xiro}
\end{equation}%
and
\begin{equation}
\xi _{T}^{\ast }=\frac{B_{pT}\left( \alpha +q\right) -2B_{p}\left( \alpha
_{T}+q_{T}\right) }{4\sqrt{B_{p}}N_{0}\rho \left( \alpha +q\right) ^{2}}+ \\
\frac{2N_{0}\rho \sqrt{B_{p}}\left[ \alpha q_{T}-q\alpha _{T}+\bar{\varsigma}%
\left( \alpha _{T}+q_{T}\right) \right] }{4\;\sqrt{B_{p}}N_{0}\;\rho \left(
\alpha +q\right) ^{2}},  \label{xiT}
\end{equation}%
here $B_{pT}=\partial B_{p}/\partial T$, $B_{p\rho }=\partial B_{p}/\partial
\rho $, $\alpha _{T}=\partial \alpha /\partial T$ and $q_{T}=\partial
q/\partial T$. Similarly to the previous sub-section, from Eqs. (\ref{k0}), (%
\ref{Ein0}), and (\ref{Q}) one may calculate both, the real and imaginary
parts of $c$ and $k$. However, for this particular plasma $\xi ^{\ast }$ is
a function of both $\rho $, and $T$ instead of $T$ only as given by
Eq. (\ref{xi0C}).

Fig.(2a) is a $3D$ plot of the ionization rate $\xi^{\ast}$ as function of $
T(K)$ and density $n(cm^{-3})$ (in Fig. (2a) the red color refers to
temperatures close to 5.000 $K$; on the other hand, the magenta color gives
the highest temperatures, which are of the order of 30.000 $K$, for a fixed
value of the photo-ionization $\varsigma = 5 \times 10^{-13} s^{-1}$).
Fig.(2b) shows a $3D$ plot of the ionization $\xi^{\ast}$ as a function of
the density $n(cm^{-3})$ and the photo-ionization $\varsigma(s^{-1})$, for a
fixed value of temperature ($log T = 4.16 K$), spanning in the range of
values for the galactic interstellar medium \cite{KG14}. In Fig. (2b) the
color indicates the values of the density, red color refers to densities $n$
near zero values, and magenta color indicates values of the density $n$
close to 100 $(cm^{-3}$). From both Figs. (2a), and (2b) follows that the
effect of the ionizing radiation is to increase the ionization at any
temperature, respect to that resulting by collisions only, however the
strong ionization occurring at temperature $\approx 2 \times 10^{4}$ K is
determined by collisions, for galactic values of the photo-ionization rate $%
\varsigma(s^{-1})$.

As it can be verify the presence of the ionizing radiation field shifts the
value of $\xi_{T}^{\ast}$ towards higher temperatures ($\log T = 4.21$, for $%
\varsigma= 5 \times 10^{-13} s^{-1}$) and smooth the change of the damping
per unit wave length with the temperature for any wave frequency, as it can
be shown comparing Fig. (1a) with Fig. (3a) in which the damping $2 \; \pi
\; k_{i}/k_{r}$ is plotted as a function of $T$, for $\varsigma= 5 \times
10^{-13} s^{-1}$, and the same three values of $\omega \tau$ shown in
Fig. (1a) but for a rate given by (\ref{CFX}) instead of (\ref{XCol}). The
change of value of the maxima of the damping per unit wave depends of the
value of $\omega \tau$, in particular increases for $\omega \; \tau= 1$,
additionally they are shifted towards higher values of $T$ following the
shift of the maximum of $\xi_{T}^{\ast}$ as follows from physical
considerations.

Accordingly, the change of the phase velocity produced (taking into account
the photo-ionization) can be seen comparing Fig. 3b with Fig. 1b. The
minimum is shifted towards higher temperatures but they are smoothed at high
frequencies ($\omega \tau$) as shown by comparing the point lines ($
\omega \tau=10$) in the the above two figures.

At a particular temperature, the changes of the damping per unit wave length
and the corresponding to the phase velocity are small (for galactic vales of
the photo-ionization $\varsigma$) as it can be shown comparing Figs. (3c)
with (1c) and Figs. (3d) with (1d), respectively. Generally, the qualitative
and quantitative effects of the photo-ionization are small respect to those
produced by collisions only in an atomic Hydrogen gas, as far as sound wave
propagation is concerned, and in the range of values of the parameters above
considered.

\subsection{Physical Implications}

The aim of the present section is to compare the value of the three
absorption coefficients corresponding to: (1) the bulk viscosity $\tilde{k}%
_{b} =c_{T} k_{i}/\omega$, (2) the dynamical viscosity $\tilde{k}_{\nu}$,
and (3) the thermal conduction $\tilde{k}_{\kappa}$ which are given by %
\citep{LL81,LL86}, i.e.
\begin{equation}
\tilde{k}_{\nu}=\frac{2 \; \omega \; \nu}{3 \; c_{T}^{2} \; \gamma^{3/2}},
\label{dy}
\end{equation}
where $\nu$ is the kinematic viscosity, and
\begin{equation}
\tilde{k}_{\kappa}=\frac{\omega\;\left(\gamma-1\right) \;\chi}{%
2\;c_{T}^{2}\;\gamma^{3/2}},  \label{cond}
\end{equation}
in which $\chi$ corresponds to the thermometric conductivity, %
\citep{Parker57,Sp62,Bra65,LL81}. The problem of sound wave propagation in a
self-consistent model of the atomic gas in the galaxy and other plasmas of
interest in Astrophysics, for which $H_e$ and ions of $H_e$, and ions of
heavy elements included will be published elsewhere.

Incidentally, another irreversible process in plasmas, is due to the
frictional force between ions of mass $m_{i}$ (and velocity $\mathbf{\ v}%
_{i} $), and neutral particles of mass $m_{n}$ (and velocity $\mathbf{v}_{n}$%
) \citep{Bra65}. The time scale for equalizing the velocities can be easily
calculated from the respective \emph{Braginskii relations}, from which one
obtains the equation
\begin{equation}
\tau_{ni}\approx\frac{(m_{i}+m_{n})}{\left\langle \sigma v\right\rangle
(\rho_{i}+\rho_{n})},  \label{Bragi}
\end{equation}
where $\left\langle \sigma v\right\rangle$ is a mean value of the product of
the cross-section and the relative velocity averaged over all velocities. As
it can be easily verified, generally $\tau_{ni} << \tau$, additionally the
frictional damping becomes independent on the wave-length $\lambda$, and it
is only important for oscillations with very high frequencies, and in
plasmas with very low ionization \citep{Bra65,NKM99,W04}. Such effect will
not be considered at the present discussion.

Figs. (4a), (4b), and (4c), are plots of the absorption coefficients $k_{b}$
(thin line), $k_{\nu}$(dash line), $k_{\kappa}$ (point line), and the total
absorption $k_{tot}$ $=k_{b}+$ $k_{\nu}+$ $k_{\kappa}$ (thick line) in units
of $cm^{-1}$, as functions of the temperature $T$ for $n=1$ $(cm^{-3})$, a
photo-ionization rate value of $\varsigma=5\times10^{-13}$s$^{-1}$, and
three different values of the frequency $\omega \; \tau=10^{-1}$, $1$ and $%
10 $, respectively. The relaxation time is plotted in Fig. (4d) for $n=1$
and three values of the photo-ionization rate $\varsigma = 5 \times10^{-14}$
(dash line), $5\times 10^{-13}$ (thick line) and $10^{-12}$\ (point line), $%
s^{-1}$. Due to the fact that the effect of damping of sound waves is
linear, it is worthy to calculate the total absorption coefficients due to
the above three effects. The absorption by bulk viscosity becomes the
dominant one in the range of temperatures where recombination-ionization
takes place $4.2 \times 10^{3} \lesssim T \lesssim T_{M}(\omega \tau)$,
where $T_{M}(\omega \tau)$ is a function of the wave frequency,
increasing when $\omega \tau$ decreases as can be seen in the above Figs.
(4a), (4b), and (4c). At high temperatures ($T > T_{M}$) and high
ionization, the thermal conduction (by electrons) dominates, instead, at low
temperatures $T \lesssim 4.2 \times 10^{3}$ $K$, the thermal conduction by
neutral atoms becomes the dominating one. At frequencies $\omega \tau
\gtrsim 1$ the bulk viscosity coefficient shows a conspicuous (relative)
maximum. On the other hand dynamical viscosity is much more lower (more than
one order of magnitude) than both, bulk viscosity, and thermal conduction in
the range of temperature under consideration. In conclusion, in a
photo-ionized Hydrogen plasma the bulk viscosity is the most important
damping mechanism in the range of temperature $4.2 \times 10^{3} \lesssim T
\lesssim T_{M}(\omega \tau)$. Fig.(4d) is a plot of the relaxation time $%
\tau$ ($=|X_{\xi}|^{-1}$ $s$) as a function of temperature for three
different values of the photo-ionization rate $\varsigma=5\times10^{-14}$%
(dash line), $\varsigma=5\times10^{-13}$ (thick line), $\varsigma=10^{-12}$
(point line). As it is expected the relaxation time $\tau$ sharply decreases
at $T \sim 10^{4} \sim 2$ \bigskip $\times 10^{4}K$ and its value is close
to $\tau \sim 10^{5}$ $y$. Therefore, for a typical number density $n \sim 1$
(for instance) the space scale for damping ranges between $\sim 0.03$ $pc$
at high temperatures ($\sim 5 \times 10^{4} K$), and $\sim 30$ $pc$ at low
temperatures ($\sim 10^{4} K$), for frequencies $\omega \; \sim 10^{-5}
y^{-1}$.

In Summary, following the Einstein (1920) \citep{Ein} work based on propagation of sound
waves in reacting gases, the bulk viscosity coefficient introduced by %
Landau \& Lifshitz \citep{LL86} (Eq .\ref{LLvis2}) has been generalized to chemically active
gases. The bulk viscosity coefficient $\sim k_{i}$ becomes the imaginary
part of the wave vector $k$ calculated from Eqs.(\ref{k0}, \ref{ccol} and
 \ref{QEins}). In particular, for a collisionally ionized Hydrogen gas, the
bulk viscosity in the Landau approximation becomes zero, but it is different
from zero at the present approximation, see results in section \ref{appl}.
For context, additionally the bulk viscosity is also calculated for a
photo-ionized Hydrogen gas for values of parameters characteristic of the
high latitude atomic gas in the Galaxy.

\newpage

\begin{figure}[ptb]
\begin{center}
\includegraphics[width = 6.0 in, height= 5.0 in]{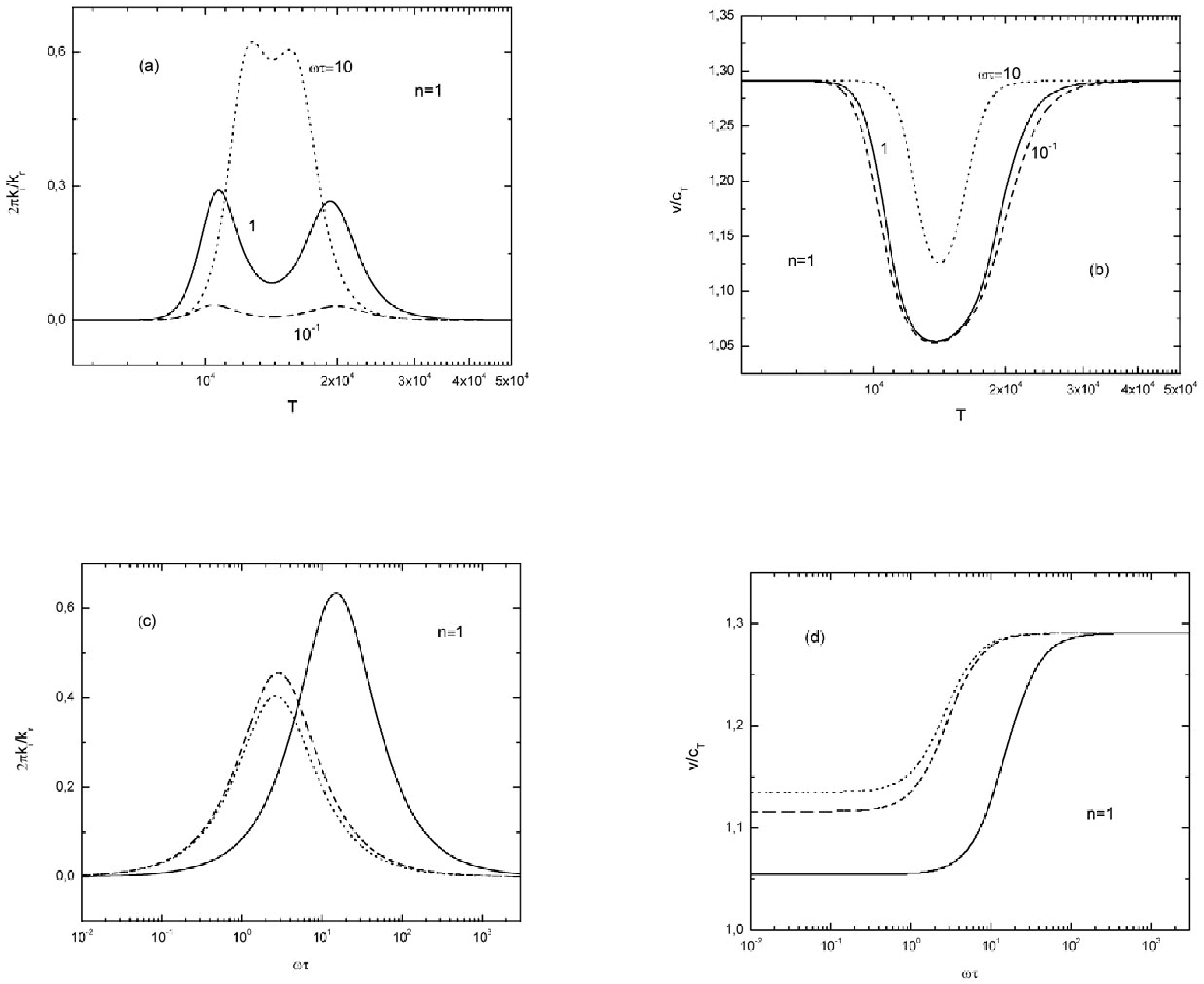}
\end{center}
\caption{\textbf{1a.} The damping per unit wave length $2 \protect\pi %
k_{i}/k_{r}$ as a function of temperature for three different values of the
dimensionless frequency $\protect\omega \protect\tau =10^{-1}$ (dash line), $
1$ (thick line ) and $10$ (point line). \textbf{1b.} The phase velocity $
v_{ph}/c_{T}$ normalized to the isothermal sound speed $c_{T}$ ($= \protect%
\sqrt{p_{0}/\protect\rho _{0}}$) as a function of temperature for three
different values of the dimensionless frequency $\protect\omega \protect\tau %
= 10^{-1}$ (dash line), $1$ (thick line ) and $10$ (point line). \textbf{1c.}
The damping per unit wave length $2\protect\pi k_{i}/k_{r}$ as a function of
the dimensionless frequency for three different values of the temperature $
\log T=4.04$ (dash line), $\log T=4.16$ (thick line) and $\log T=4.28$,
point lines). \textbf{1d.} The phase velocity $v_{ph}/c_{T}$ normalized to
the isothermal sound speed as a function of the dimensionless frequency for
three different values of the temperature $\log T=4.04$ (dash line), $\log
T=4.16$ (thick line) and $\log T=4.28$, (point lines).}
\label{1}
\end{figure}

\begin{figure}[ptb]
\includegraphics[width = 6.0 in, height= 5.5 in]{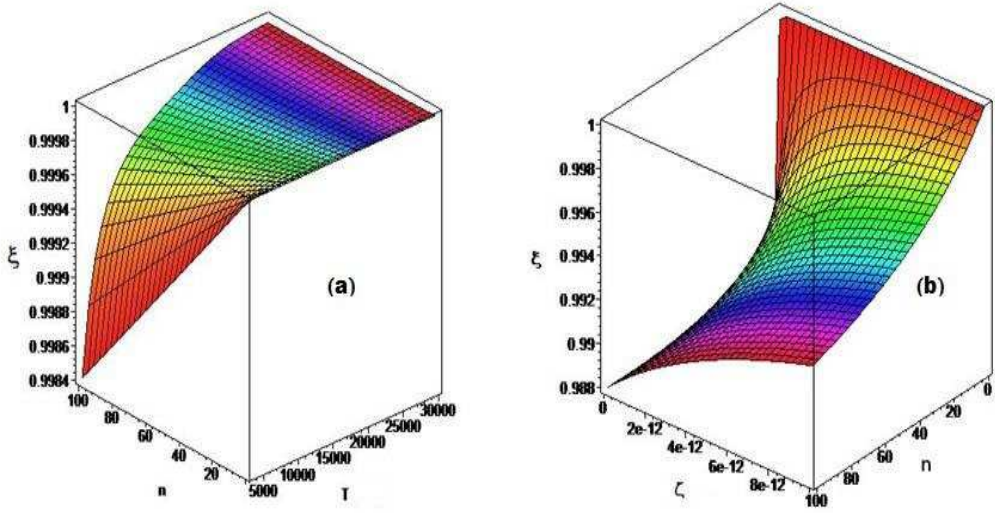}
\caption{\textbf{2a.} The equilibrium ionization $\protect\xi^{\ast}$ as a
function of temperature $T(K)$ and density $n$ (cm$^{-3}$) for a
photo-ionization rate $\protect\varsigma=$ $5\times 10^{-13} $s$^{-1}$ (In
Fig. \textbf{2a}, the red color refers to temperatures close to 5.000 $K$,
and the magenta color refers to the highest temperatures of the order of the
30.000 $K$). \textbf{2b.} The equilibrium ionization $\protect\xi^{\ast}$ in
a $3D$ plot as a function of the density $n(cm^{-3})$ and the ionization
rate $\protect\varsigma(s^{-1})$ for a fixed value of temperature ($log \; T
= 4.16 K$), and where the color palette indicates the values of the density,
i.e. red color refers to $n$ around zero, and magenta color indicates values
of the density $n$ close to 100 $(cm^{-3}$).}
\label{2}
\end{figure}

\begin{figure}[ptb]
\includegraphics[width = 6.0 in, height= 5.5 in]{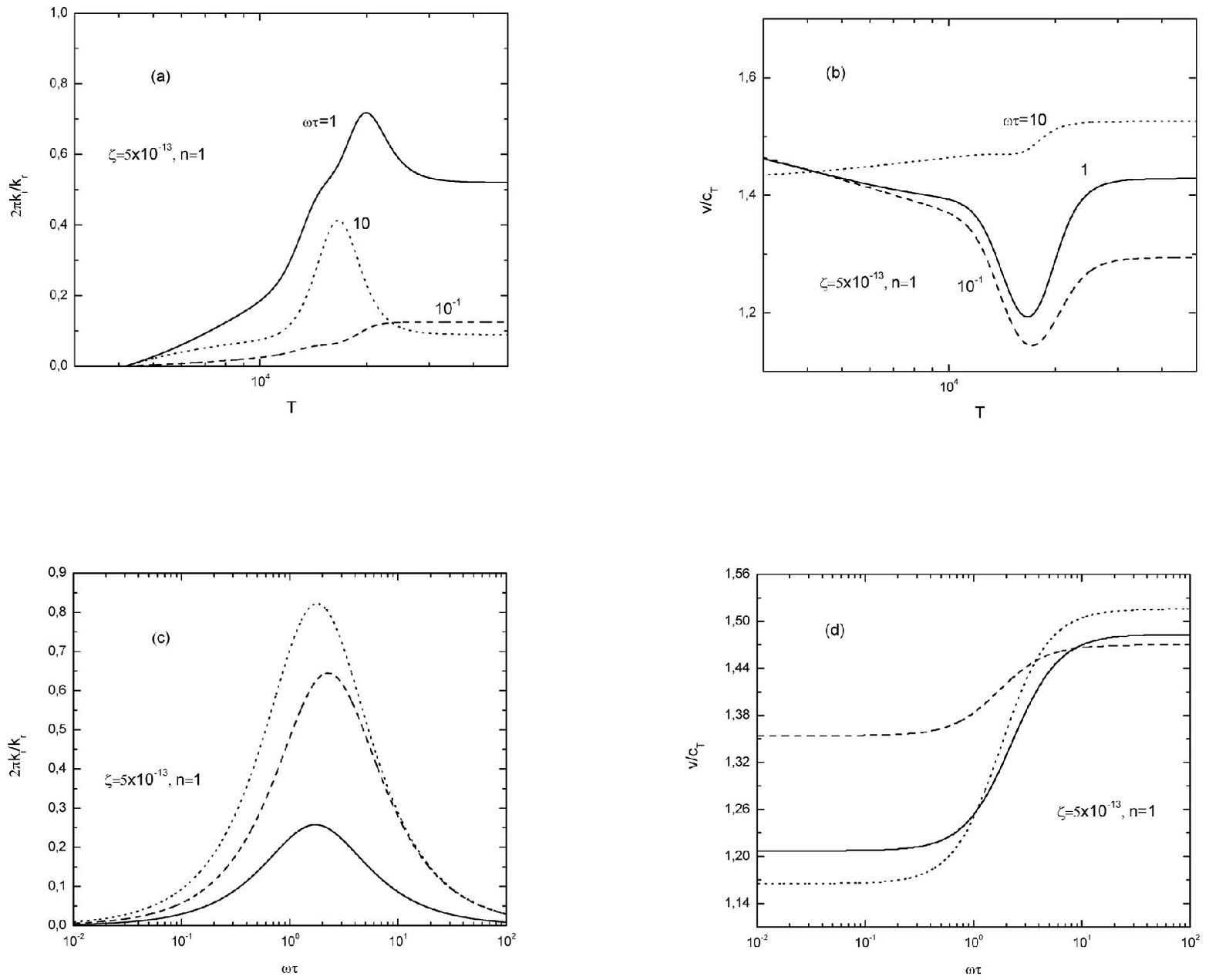}
\caption{\textbf{3a.} As Fig.\textbf{(1a)} for a photo-ionized gas with a
rate given by the expression (\protect\ref{CFX}) and $\protect\varsigma=
5\times10^{-13} $s$^{-1}$. \textbf{3b.} As Fig.\textbf{(1b)} for a
photo-ionized gas with a rate given by the expression (\protect\ref{CFX})
and $\protect\varsigma= 5 \times10^{-13} $s$^{-1}$. \textbf{3c.} As Fig.
\textbf{(1c)} for $\protect\varsigma= 5\times10^{-13} $s$^{-1}$. \textbf{3d.}
As Fig. \textbf{(1d)} for $\protect\varsigma= 5\times10^{-13} $s$^{-1}$.}
\label{3}
\end{figure}

\begin{figure}[ptb]
\includegraphics[width = 6.4 in, height= 5.5 in]{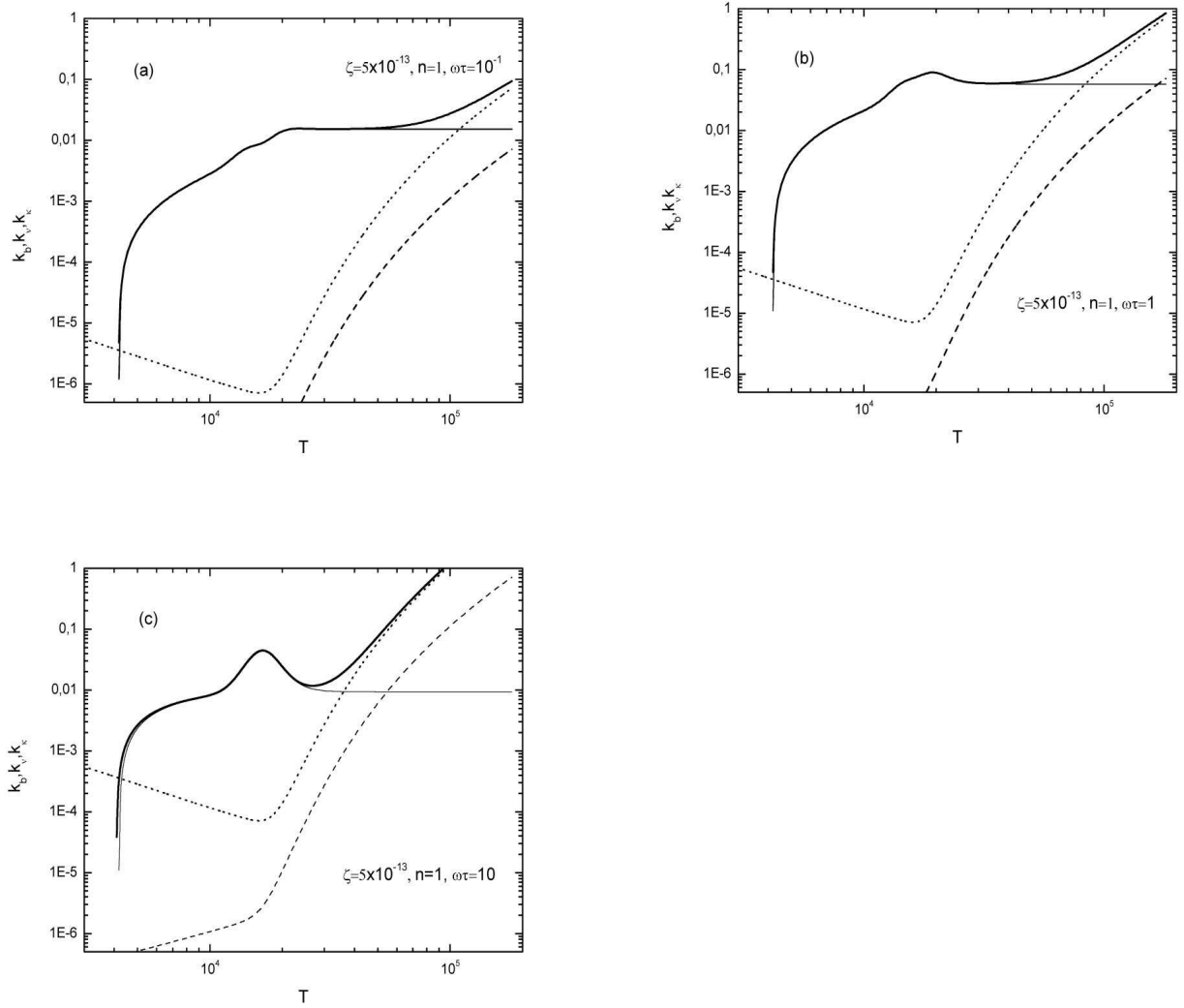}
\caption{\textbf{4a.} The absorption coefficients ${k}_{b}$ (thin line), $_{%
\protect\nu}$(dash line), ${k}_{\protect\kappa}$ (point line) and the total
absorption (thick line) ${k}_{tot}={k}_{b}+$ ${k}_{\protect\nu}$ $+{k}_{%
\protect\kappa}$ as functions of temperature $T$ for $n=1$ cm$^{-3}$, a
photo-ionization rate $\protect\varsigma=5\times10^{-13}$s$^{-1}$ and a
dimensionless frequency $\protect\omega\protect\tau=10^{-1}$. \textbf{4b.}
As Fig. \textbf{(4a)} for $\protect\omega\protect\tau=1$. \textbf{4c.} as
Fig.\textbf{(4a)} for $\protect\omega\protect\tau=10$. \textbf{4d.} is a
plot of the relaxation time $\protect\tau$ ($=|X_{\protect\xi}|^{-1}$ $(s)$)
as a function of temperature $T$ for three different values of the
photo-ionization rate $\protect\varsigma=5\times10^{-14}$(dash line), $
\protect\varsigma=5\times10^{-13}$ (thick line), and $\protect\varsigma %
=10^{-12}$ (point line)}
\label{4}
\end{figure}

\end{document}